\documentclass{ragtime} 

\title[Causal horizons and topics in structure formation]
{Causal horizons and some topics concerning structure formation}

\author[J.C. Miller and I. Musco]
{John Miller\at[]{1,2,a} and Ilia Musco\at[]{3,4}\\
\ins{1}SISSA, International School for Advanced Studies, Via Bonomea 
265,\splitins[1] 
I-34136 Trieste, Italy\\
\ins{2}Department of Physics (Astrophysics), University of Oxford, 
Keble Road,\splitins[2] 
Oxford OX1 3RH, UK\\
\ins{3}Centre of Mathematics for Applications, Department of 
Mathematics,\splitins[3]
University of Oslo, PO Box 1053 Blindern, NO-0316 Oslo, Norway\\
\ins{4}Laboratoire Univers et Th\'eories, UMR 8102 CNRS, Observatoire de 
Paris,\splitins[4]
Universit\'e Paris Diderot, 5 Place Jules Janssen, F-92190 Meudon, France\\ 
\ins{a}\Email{jcm@astro.ox.ac.uk}}

\providecommand{\dif}{\mathrm{d}} 

\begin{document}

\begin{abstract} 

This is a write-up of a talk given at the Opava RAGtime meeting in 2011, but it 
has been updated to include some subsequent related developments. The talk 
focused on discussion of some aspects of black hole and cosmological horizons 
under rather general circumstances, and on two different topics related to 
formation of cosmological structures at different epochs of the universe: 
virialization of cold dark matter during standard structure formation in the 
matter-dominated era, and primordial black hole formation during the radiative 
era.

\end{abstract}

\begin{keywords}
black hole physics~-- early universe~-- large-scale structure of the universe
\end{keywords}

\section{Introduction}

This presentation focuses firstly on two different types of causal horizon: 
those for black holes (where no causal signal can get {\em out} from inside), 
and that for the universe (where no causal signal can get {\em in} from 
outside). Also, we discuss some topics connected with formation of structure 
in the universe in the matter-dominated and radiative eras. We follow the 
convention of using units for which $c=G=1$ except in section 
\ref{matter-dominated}, where the treatment is entirely Newtonian and it is 
convenient to retain $G$.

In all of these discussions, we will make the (major) simplification of 
considering just spherical symmetry but, apart from that, we will remain 
rather general. We start from the Friedman-Robertson-Walker metric for 
describing a homogeneous and isotropic background universe and we use the 
spatially-flat form of it, in line with current observations:
 \begin{equation}
\dif s^2 = - \dif t^2 + S^2(t)\left[\dif r^2 + r^2\left(\dif \theta^2 + 
\sin^2\theta \dif \varphi^2\right)\right] \, ,  
\end{equation}
 where $r$ is a co-moving radial coordinate and $S(t)$ is the scale factor. 
This can be written in the alternative form 
 \begin{equation}
\dif s^2 = - \dif t^2 + S^2(t)\dif r^2 + R^2\left(\dif \theta^2 + 
\sin^2\theta \dif \varphi^2\right) \, , 
\end{equation}
 where $R = S(t)r$ is a circumference coordinate (invariantly defined as 
being the proper circumference of a circle, centred on the origin, divided by 
$2\pi$). This is the same quantity as used for the radial coordinate in the 
standard form of the Schwarzschild metric.

The above description is for a uniform medium; when we have a (spherically 
symmetric) deviation away from this, the metric can then be written in the 
generalised form
 \begin{equation}
\dif s^2 = - a^2 \dif t^2 + b^2 \dif r^2 + R^2\left(\dif \theta^2 + 
\sin^2\theta \dif \varphi^2\right) \, ,
\label{cosmic_time}
\end{equation}
 with $a$, $b$ and $R$ all being functions of $r$ and $t$. Using a diagonal 
form of the metric like this (with no cross terms involving $\dif r\,\dif t$, 
etc.) implies a particular choice of time slicing, and the time coordinate 
here is often called ``cosmic time''. The form of metric (\ref{cosmic_time}) 
can be used in principle for any spherically-symmetric space-time, although 
it is often more convenient in practice to use other kinds of slicing.

\section{Causal horizons}

In this section, we discuss how the concepts of black-hole and cosmological 
horizons emerge from a general treatment of outgoing and ingoing null rays. 
We continue to assume spherical symmetry and take the medium to be a perfect 
fluid, but our discussion is general in the sense that it is independent of 
the equation of state used for the matter and we make no assumptions of 
homogeneity (with reference to the cosmological case), or of stationarity, 
asymptotic flatness and the presence of vacuum exteriors (with reference to 
the black holes). It can be interesting to see how well-known results emerge 
in this approach.

First, we consider the general treatment of radial null rays, using the 
cosmic time form of the metric (\ref{cosmic_time}) introduced above. Along 
the path of any radial null ray, we have $\dif s = \dif \theta = \dif \phi = 
0$ and so
 \begin{equation}
\dif r = \pm\, \frac{a}{b}\,\dif t \, ,
\end{equation}
 with the plus corresponding to an outgoing ray and the minus to an ingoing 
one. Note that here ``outgoing'' and ``ingoing'' are defined with respect to 
the comoving frame of local matter. This convention is used throughout the 
present section.

The general expression for changes in $R$ along a radial worldline is
 \begin{equation}
\dif R = \frac{\partial R}{\partial t}\,\dif t + \frac{\partial R}{\partial 
r}\,\dif r \, ,
\end{equation}
 and so along a radial {\em null ray}
 \begin{equation}
\dif R = \left(\frac{\partial R}{\partial t} \pm \frac{a}{b} 
\frac{\partial R}{\partial r}\right) \dif t \, .
\label{RNR2}
\end{equation}

Following the classic paper of \citet{misner69}, we now introduce the 
operators
 \begin{equation}
D_t \equiv \frac{1}{a}\,\frac{\partial}{\partial t} \qquad {\rm and} 
\qquad D_r \equiv \frac{1}{b}\,\frac{\partial}{\partial r} \, .
\end{equation}
 Applying these to the circumference coordinate $R$, one then defines the 
quantities 
 \begin{equation} 
U \equiv D_t R \qquad {\rm and} \qquad \Gamma \equiv D_r R \, , 
\end{equation}
 where $U$ is the radial component of four-velocity in the ``Eulerian'' 
frame, with respect to which the fluid is moving, and $\Gamma$ is a 
generalized Lorentz factor (which reduces to the standard one in the special 
relativistic limit). In terms of these,
 \begin{equation}
\frac{\partial R}{\partial t} = aU \qquad {\rm and} \qquad 
\frac{\partial R}{\partial r} = b\Gamma \, .
\end{equation}
 Inserting these into equation (\ref{RNR2}) gives the expression for how $R$ 
changes with time along a radial null ray: 
 \begin{equation}
\frac{\dif R}{\dif t} = a \left(U \pm \Gamma \right) \, ,
\label{null_ray}
\end{equation}
 where the plus is again for a ray which is outgoing (with respect to the 
matter) while the minus is for an ingoing one.

To find an expression for $\Gamma$, we need to use the Einstein field 
equation. As usual, we approximate the matter to behave as a perfect fluid 
with the stress-energy tensor
 \begin{equation}
T^{\mu\nu} = (e+p) u^\mu u^\nu + p g^{\mu\nu} \, ,
\end{equation}
 where $e$ is the energy density, $p$ is the pressure and $u^\mu$ is the 
four-velocity. The $G^0_0$ and $G^1_1$ components of the Einstein equation 
then give
 \begin{equation}
4\pi R^2 e R_{,r} = \textstyle{\frac{1}{2}}\left(R+RU^2-R\Gamma^2\right)_{,r} 
\, ,
\label{G00}
\end{equation}
 and
 \begin{equation}
4\pi R^2 apU = -\textstyle{\frac{1}{2}}\left(R+RU^2-R\Gamma^2\right)_{,t} 
\, ,
\label{G11}
\end{equation}
 with the commas representing partial derivatives. It is convenient to make 
the definition
 \begin{equation}
m \equiv \textstyle{\frac{1}{2}}\left(R+RU^2-R\Gamma^2\right)\, .
\label{m}
\end{equation}
 Integrating equation (\ref{G00}) then gives
 \begin{equation}
m = \int 4\pi R^2 e\,\dif R \, ,
\end{equation}
 (corresponding to the interpretation of $m$ as the mass contained within 
radius $R$), while equation (\ref{G11}) gives 
 \begin{equation}
D_t m=-4\pi R^2 pU \, ,
\end{equation}
 (representing the change of energy resulting from work done against pressure 
during expansion or contraction). Rearranging the terms in (\ref{m}) then 
gives
 \begin{equation}
\Gamma^2=1+U^2-\frac{2m}{R}\, .
\label{Gamma2}
\end{equation}

Returning now to equation (\ref{null_ray}), the limiting surface at which an 
outgoing radial light ray cannot move to larger $R$, is given by
 \begin{equation}
\left(\frac{\dif R}{\dif t}\right)_{\rm \! out} = a (U + \Gamma) = 0 \, ,
\end{equation}
 implying
 \begin{equation}
\Gamma = - U \, .
\label{trap1}
\end{equation}
 This corresponds to the so-called ``apparent horizon'' of a black hole. 
Similarly, the limiting surface at which an ingoing radial light ray cannot 
move to smaller $R$ is given by
 \begin{equation}
\left(\frac{\dif R}{\dif t}\right)_{\rm \!in} = a (U - \Gamma) = 0 \, ,
\end{equation}
 implying
 \begin{equation}
\Gamma = U \, .
\label{trap2}
\end{equation}
 This corresponds to the cosmological (Hubble) horizon. Note that the 
surfaces for which (\ref{trap1}) and (\ref{trap2}) hold are {\em marginally 
trapped surfaces} and so are representations of a concept \citep{penrose65} 
which plays a fundamental role in general relativity.

Conditions (\ref{trap1}) and (\ref{trap2}) are different, of course, but for 
both of them
 \begin{equation}
\Gamma^2 = U^2 \, ,
\end{equation}
 and so, using (\ref{Gamma2}), they both correspond to the condition
 \begin{equation}
R = 2m \, ,
\end{equation}
 which is a familiar result! Although we have used cosmic-time slicing in 
this derivation, the final result is actually independent of the slicing 
used. We stress again that our derivation here does not depend on any 
assumptions of homogeneity (with reference to the cosmological case), or of 
stationarity, asymptotic flatness and the presence of vacuum exteriors (with 
reference to black holes).

\section{Cosmological structure formation}\label{structure}

In this section, we discuss some topics concerning cosmological structure 
formation at two different stages in the history of the universe: in the 
matter-dominated and radiative eras. The main interest is in the consequences of 
perturbations which started as small quantum fluctuations in the very early 
universe and were then inflated onto supra-horizon scales, eventually 
re-entering the horizon as the universe continued to expand, and becoming 
causally connected again. (``Horizon'' here refers to the {\em cosmological} 
horizon.) We focus on the case of an over-density surrounded by a compensating 
under-density. Once the over-density has re-entered the horizon, there is a 
possibility that it could then evolve into a persisting condensed structure. A 
perturbation originating at a very early time, such as those mentioned above, 
may have begun with a mixture of growing and decaying components, but any 
decaying part would soon have faded away so that by the time of horizon 
re-entry, only the growing part would remain. Growing-modes are special; they 
have a particular combination of density and velocity perturbation which makes 
them ``hold together'' as they evolve.

We discuss below the re-entry of perturbations during the {\em matter-dominated 
era}, when the matter is commonly described as pressureless, with $p = 0$ 
(although we will have more to say about that), and during those parts of the 
{\em radiative era} (defined as being when only relativistic zero-rest-mass 
particles are important) in which $p = \frac{1}{3}e$ is a good approximation. 
The first case can lead to formation of galactic or pre-galactic equilibrium 
structures, whereas in the second case primordial black holes (on a much 
smaller scale) are the only condensed structures that can be formed.

\subsection{Virialization in the matter-dominated era}\label{matter-dominated}

Cosmologists like to use equations of state of the form $p = we$, where $w$ 
is a constant. The cases $p = \frac{1}{3}e$ and $p = 0$ do fit with this, of 
course, but it is questionable whether taking $p = 0$ actually makes sense in 
general for the matter-dominated era. For calculating the evolution of a 
uniform background universe, it is indeed satisfactory, but it becomes 
problematic when dealing with structure formation beyond the regime of linear 
perturbations. For cold dark matter (CDM) particles, it is frequently said 
that they must be pressureless because of being effectively collisionless, 
but this misses the point that pressure comes from the random motion of 
particles and is only indirectly influenced by collisions between them. If 
CDM particles have a non-zero velocity dispersion, then they automatically 
have a non-zero pressure and this is generally not irrelevant even if it may 
be small. The role of collisions is in assisting the particle distribution 
function to relax towards an isotropic Maxwellian, not directly in producing 
the pressure. A completely collisionless medium can certainly have a finite 
pressure (although that will generally not be isotropic).

In this subsection, we investigate the phenomenology of the ``turn-round 
radius'' and the ``virialization radius'' for perturbations when they 
re-enter the cosmological horizon and begin to feel their self-gravity. 
Initially, the over-density is continuing to expand along with the rest of 
the universe (although slightly more slowly because of the velocity 
perturbation in a growing mode) but, as it progressively begins to feel its 
self-gravity more, it slows down further and eventually reverses its 
expansion into a contraction. Its radius when that happens is called the 
turn-round radius. We will follow here just the subsequent behaviour of the 
dark matter component, which is more or less collisionless. As the 
contraction proceeds, the random velocities of the constituent particles 
progressively increase until eventually the effect of their random motions is 
sufficient to balance gravity (possibly aided by rotation) and the 
configuration settles into an equilibrium state. Its radius then is called 
the virialization radius (we explain this more below). In numerical 
simulations, it is often found that this virialization radius is roughly half 
of the turn-round radius.

Clearly, structure formation is in general a three-dimensional problem, but 
could one get a reasonable approximate picture for the above process by using 
a simple spherically-symmetric toy model? If so, that could be useful for 
trial inclusion of further effects (dynamical scalar fields, etc.). Our idea 
is to include the random motions of the CDM particles in terms of an 
effective temperature $T$ and insert that into a model equation of state, 
giving a pressure $p$. We proceed as follows. CDM particles are 
non-relativistic and so the thermal energy per particle is given by
 \begin{equation}
u = \textstyle{\frac{3}{2}}\, k_B T \, ,
\end{equation}
 (assuming local isotropy; $k_B$ is Boltzmann's constant). The thermal energy 
density is then
 \begin{equation}
\rho\varepsilon = \textstyle{\frac{3}{2}}\, nk_B T \, ,
\end{equation}
 (where $\rho$ is the rest-mass density, $\varepsilon$ is the specific 
internal energy, and $n$ is the particle number density). The ideal gas law 
$p = nk_B T$ then gives
 \begin{equation}
p = \textstyle{\frac{2}{3}}\, \rho\varepsilon \, ,
\label{EOS1}
\end{equation}
 which leads to
 \begin{equation}
p = K(s) \rho^{5/3} \, ,
\end{equation}
 using the first law of thermodynamics. This is the well-known polytropic 
relation for a monatomic non-relativistic gas (here $K(s)$ is a function of 
the specific entropy $s$ and goes to a constant for adiabatic processes).

Next, we recall the considerations leading to a simple form of the virial 
theorem, following \citet{tayler70}. (Note that the treatment in this 
subsection is entirely Newtonian and it is convenient to retain the $G$ in 
the equations for this part; also $r$ is here the standard classical radial 
coordinate.) The equation of hydrostatic equilibrium
 \begin{equation}
\frac{\dif p}{\dif r} = - \frac{Gm\rho}{r^2} \, ,
\end{equation}
 where $m$ is the mass contained within radius $r$, can be rearranged to give 
 \begin{equation}
4\pi r^3 \dif p = - \left(\frac{Gm}{r}\right) 4\pi r^2 \rho\, \dif r \, . 
\label{TOV2} 
\end{equation}
 We now integrate equation (\ref{TOV2}) over the volume of the spherical 
object:
 \begin{equation}
\int{3V\,\dif p} = \int{\Phi\,\dif m} \, ,
\end{equation}
 where $V$ is the volume contained within radius $r$ and $\Phi = 
-(Gm/r)$ is the gravitational potential at radius $r$. Integrating the 
left-hand side by parts then gives 
 \begin{equation}
3\Big[pV\Big] - 3\int{p\,\dif V} = \Omega \, ,
\end{equation}
 where $\Omega$ is the gravitational potential energy of the object. Taking 
the pressure at the surface to be zero and inserting the equation of state 
expression (\ref{EOS1}) for $p$, one obtains
 \begin{equation}
-3\int{\textstyle{\frac{2}{3}}\,\rho\varepsilon\, \dif V} = -2U = \Omega \, ,
\end{equation}
 where $U$ is the total internal energy of the object. The overall total 
energy is then 
 \begin{equation}
E = U + \Omega = \textstyle{\frac{1}{2}}\,\Omega \, ,
\label{virial}
\end{equation}
 which is negative, as it must be for a gravitationally-bound object. When 
equation (\ref{virial}) is satisfied, the configuration is said to be 
virialized.

We will now use these ideas for studying the issue of the turn-round radius 
and virialization radius within our simple toy model. We will use the 
subscripts ${tr}$ and $v$ to denote ``turn-round'' and ``virial'' 
respectively. The configuration starts off (at the turn-round radius 
$R_{tr}$) out of hydrostatic equilibrium and not satisfying equation 
(\ref{virial}). As the contraction proceeds, the pressure plays an increasing 
role in counteracting gravity until eventually equilibrium is reached and the 
virial condition (\ref{virial}) is satisfied. The radius at which this 
happens is the virialization radius ($R_v$) mentioned earlier.

Assuming conservation of the total energy during the contraction,
\begin{equation}
 E = U_v + \Omega_v = U_{tr} + \Omega_{tr} \, .
\end{equation}
 At the turn-round radius, the energy in the random motions of the CDM 
particles is still going to be small (it grows later as the contraction 
proceeds) and so it seems safe to assume that the initial total internal 
energy term $U_{tr}$ can be neglected. Doing this, and using expression 
(\ref{virial}) at the virial radius, we have
 \begin{equation}
E = \textstyle{\frac{1}{2}}\,\Omega_v = \Omega_{tr} \, .
\end{equation}
 If we now make the (rough) assumptions that the total mass $M$ does not 
change during the contraction and that, throughout, we can write
 \begin{equation}
\Omega = - \frac{GM^2}{R} \times {\rm constant} \, ,
\end{equation}
 then $\frac{1}{2}\Omega_v = \Omega_{tr}$ gives 
 \begin{equation}
\frac{GM^2}{2R_v} = \frac{GM^2}{R_{tr}} \, ,
\end{equation}
 and so
 \begin{equation}
R_v = \textstyle{\frac{1}{2}}R_{tr} \, ,
\end{equation}
 in agreement with the numerical results.

Our purpose here has been to suggest that this type of ``fluid'' treatment of 
cold dark matter might be a useful approach in some circumstances. Clearly, 
implementations could be made much more detailed than the one which we have 
sketched above.

\subsection{The radiative era and primordial black holes}
 
\begin{figure}[t!]
\centering
\includegraphics[clip=true,bb=154 560 336 746,width=7cm]{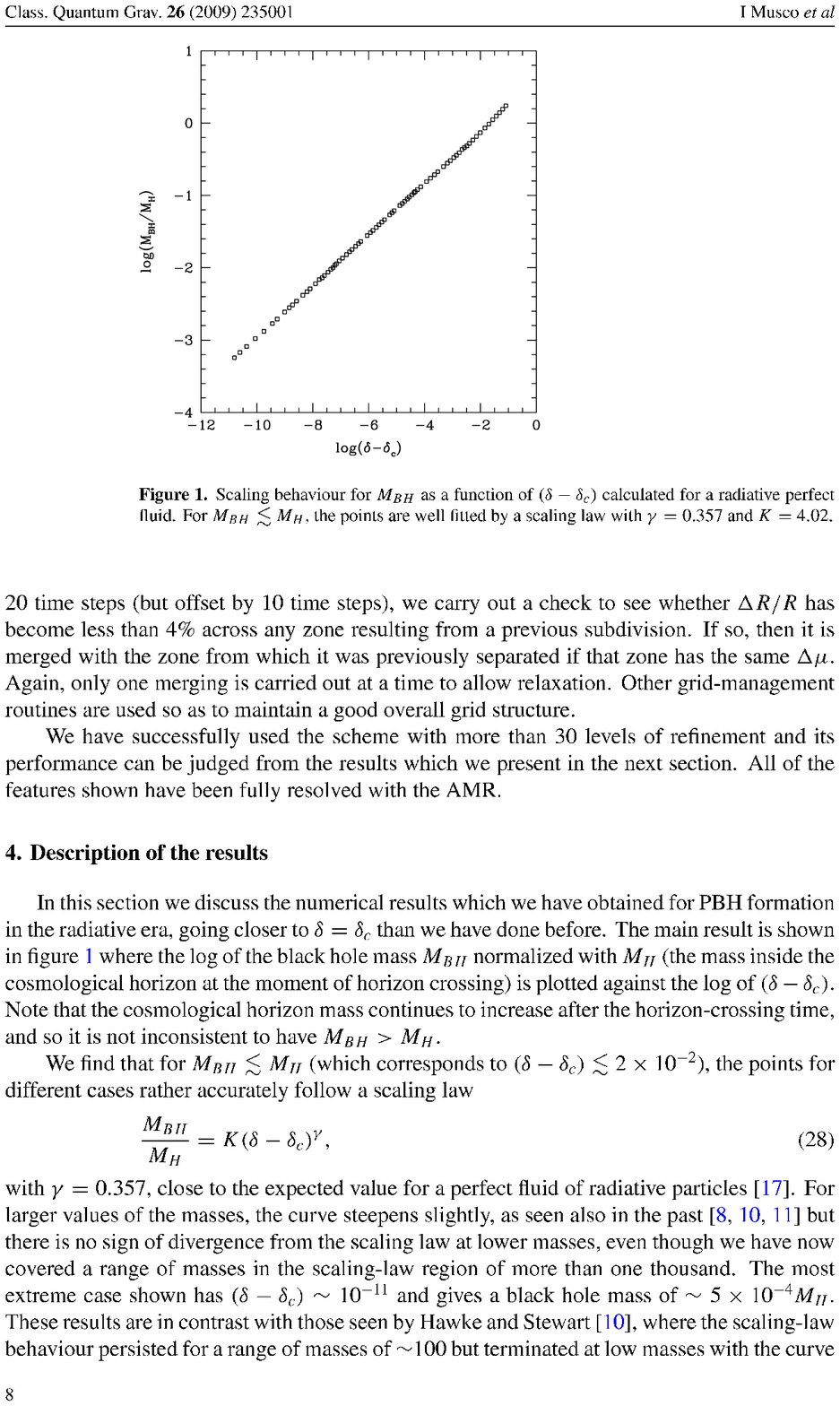}   
 \caption{\label{fig.1} \small Scaling behaviour for $M_{BH}$ as a function 
of $(\delta-\delta_c)$. $M_{BH}$ is measured in units of $M_H$, the mass 
within the cosmological horizon when the perturbation re-enters it. For 
$M_{BH} \lesssim M_H$, the points are well fitted by a scaling law with 
$\gamma = 0.357$, which matches well with the corresponding result obtained 
semi-analytically by \citet{maison96} for standard critical collapse with 
this type of matter.}
 \end{figure}

Objects composed of matter for which $p = \frac{1}{3}e$ have an adiabatic index 
of $4/3$ and are fundamentally unstable. Because of this, collapsing 
perturbations in the radiative era do not produce equilibrium condensed 
structures, but either form black holes (if the perturbation amplitude $\delta$ 
is greater than a certain critical threshold value $\delta_c$) or bounce and 
return back into the roughly uniform medium from which they came. Black holes 
formed then could have lower masses than ones formed today by the collapse of 
stars. Since this type of matter has no intrinsic scale, the question arises of 
whether the phenomenon known as ``critical collapse'' \citep{choptuik93} might 
occur under these circumstances despite the background being that of the 
expanding universe. The standard form of critical collapse is characterised by 
the property that, for values of $(\delta - \delta_c)$ which are positive but 
sufficiently small, the mass of the black hole formed, $M_{BH}$, is related to 
$(\delta - \delta_c)$ by a scaling law, i.e.
 \begin{equation}
M_{BH} \propto \left(\delta-\delta_c\right)^\gamma \, ,
\label{scaling}
\end{equation}
 (with $\gamma$ being a constant) when the nature of the unperturbed 
background is kept fixed and the perturbations introduced differ in amplitude 
but not in shape. This sort of behaviour has been seen in quite a wide range 
of numerical simulations treating idealised circumstances \citep[see the 
review by][]{gundlach07} but its occurrence under ``real-world'' 
circumstances is less clear. It seemed possible that the radiative era of the 
early universe might provide an arena for this, although a potential problem 
comes from the fact that the universe itself has an intrinsic scale (the 
cosmological horizon scale) which might or might not interfere with the 
scaling behaviour. \citet{niemeyer99} made calculations which demonstrated 
the presence of a scaling law under these circumstances over a restricted 
range of $(\delta - \delta_c)$ but when more extensive calculations were 
made, going closer to the critical limit \citep{hawke02}, it was found that 
the scaling law eventually broke as the behaviour became more extreme near to 
the critical limit. We then reinvestigated this ourselves \citep{musco09}, 
focusing particularly on the use of perturbations containing only a 
growing-mode component (following on from the discussion at the beginning of 
Section \ref{structure}). For our calculations, we used a purpose-built 
numerical GR hydro code implementing an AMR technique within a null-slicing 
approach, and able to go down to extremely small values of $(\delta - 
\delta_c)$. Using growing-mode initial data, without any decaying component 
\citep[following the methodology of][]{polnarev07}, we found that the scaling 
behaviour did go all the way down to the smallest values of 
$(\delta-\delta_c)$ that we were able to treat, well beyond the breaking 
point found previously. Results are shown in Figure~\ref{fig.1}. In our work, 
we define $\delta$ as being the relative mass excess inside the over-dense 
region at the time when it re-enters the cosmological horizon, and measure 
$M_{BH}$ in units of $M_H$, the cosmological horizon mass at that time, so 
that the results are independent of epoch within the radiative era. Note 
that black holes produced like this would have typically lower masses when 
formed earlier rather than later (related with the value of $M_H$ at the 
time).

\begin{figure}[t!]
\centering
\includegraphics[width=6.55cm]{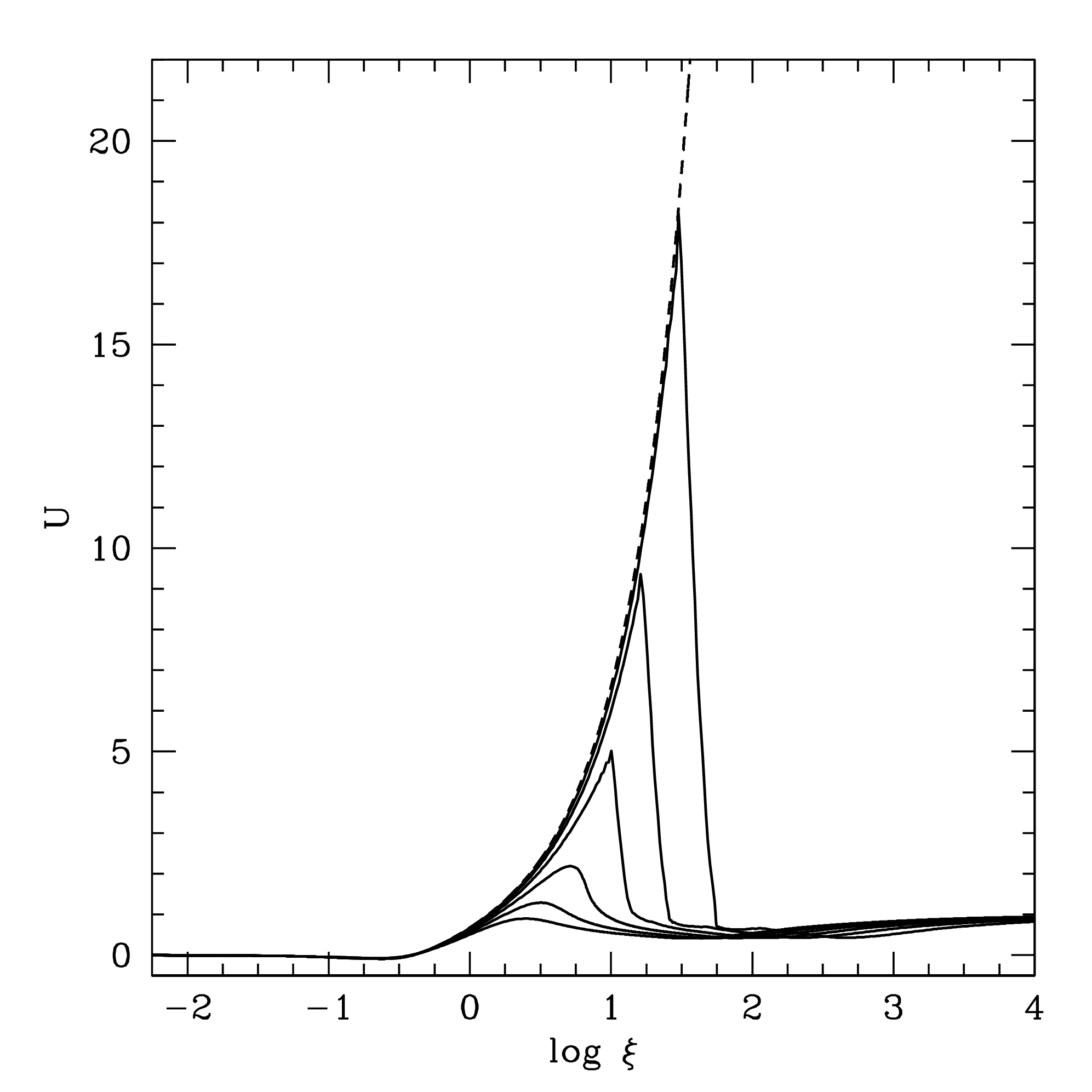} 
 \caption{\label{fig.2} \small Simulation results for the velocity $U$ (from
a run with $\delta-\delta_c \sim 10^{-9}$) plotted against the similarity 
coordinate $\xi = R/(t_c-t)$. The plot shows curves for a succession of times 
during the close approach to the similarity solution, with the higher peaks 
corresponding to the later times. The precise similarity solution for the 
collapsing matter is marked with the dashed curve, which is partly covered by 
the solid curves.}
 \end{figure}

In the literature on critical collapse, a key feature is the occurrence of 
similarity solutions accompanying the scaling laws \citep{evans94}. As 
$(\delta-\delta_c) \to 0$, a critical solution is approached where all of the 
matter in the original contracting region is progressively shed during the 
contraction which ends, with zero matter, at a time referred to as the 
critical time $t_c$. The later stages of this follow a similarity solution. 
For small positive values of $(\delta-\delta_c)$, the similarity solution is 
closely approached but eventually there is a divergence away from it, with 
the remaining material then collapsing to form a black hole. It is 
interesting to see how this plays out in our case, where the collapse occurs 
within the background of an expanding universe. We have investigated this in 
some detail \citep{musco13}. Figure~\ref{fig.2} shows our results from a run 
with $\delta-\delta_c \sim 10^{-9}$, which is rather close to the critical 
limit; the four-velocity $U$ is plotted as a function of the similarity 
coordinate $\xi = R/(t_c-t)$ at a succession of times (solid curves), with 
the higher peaks corresponding to the later times. Note the shedding of 
material occurring via a relativistic wind. One can see the progressive 
approach of the simulation results towards the similarity solution (dashed 
curve), with the range of the zone of agreement increasing with time. At the 
last time shown, the similarity solution is being closely approximated over 
all of the contracting region, where $U$ is negative (although it is quite 
hard to see this as being negative in the figure because of the scale), and 
also over the part of the surrounding region out to the maximum of $U$; 
beyond this, the simulation results diverge completely away from the 
similarity solution and eventually merge into the surrounding 
Friedmann-Robertson-Walker universe. The similarity behaviour breaks soon 
after the last time shown here, with the start of the final collapse leading 
to black hole formation. We should stress that the use of a logarithmic 
coordinate in Figure~\ref{fig.2} has the effect of making features appear 
much more abrupt than they would do with a standard linear coordinate. The 
almost vertical parts of the curves are nowhere near to being shocks and 
correspond to smoothly varying features when viewed on a linear scale.

\section{Conclusions}

We have touched here on a number of topics. Firstly, a unified treatment has 
been given of black-hole and cosmological horizons in terms of co-moving 
trapped surfaces. This does not depend on any assumptions of homogeneity (in 
the cosmological case), or of stationarity, asymptotic flatness and the 
presence of vacuum exteriors (for the black holes). We then went on to 
discuss two topics concerned with cosmological structure formation: 
virialization of cold dark matter during standard structure formation in the 
matter-dominated era, and primordial black hole formation during the 
radiative era. In the first case, we presented a simple toy model which 
serves as an analytic demonstration of phenomena observed in numerical 
simulations; in the second case, we presented results showing that black-hole 
formation by collapse of cosmological perturbations in the radiative era 
completely follows the well-known phenomenology of critical collapse, as long 
as the perturbations are of the growing-mode type when they re-enter the 
cosmological horizon.

\ack The research of Ilia Musco is currently being supported by postdoctoral
funding from ERC-StG EDECS contract no. 279954.

\bibliography{\jobname}

\end{document}